\documentclass[prd,twocolumn,showpacs,nofootinbib]{revtex4}
\usepackage{graphicx}
\usepackage[usenames]{color}
\usepackage{amsmath}
\begin{document}

\newcommand{\Caltech}{\affiliation{Theoretical Astrophysics 350-17,
    California Institute of Technology, Pasadena, California 91125, USA}}

\title{Hybrid method for understanding black-hole mergers: Head-on case}

\author{David A.\ Nichols\footnote{Electronic address: 
\texttt{davidn@tapir.caltech.edu} } } \Caltech
\author{Yanbei Chen\footnote{Electronic address: 
\texttt{yanbei@tapir.caltech.edu} } } \Caltech

\date{November 15, 2010}

\begin{abstract}
Black-hole-binary coalescence is often divided into three stages: inspiral,
merger and ringdown.
The post-Newtonian (PN) approximation treats the inspiral phase, black-hole 
perturbation (BHP) theory describes the ringdown, and the nonlinear 
dynamics of spacetime characterize the merger.
In this paper, we introduce a hybrid method that incorporates elements 
of PN and BHP theories, 
and we apply it to the head-on collision of black holes with transverse, 
anti-parallel spins.
We compare our approximation technique with a full numerical-relativity
simulation, and we find good agreement between 
the gravitational waveforms and the radiated energy and momentum.
Our results suggest that PN and BHP theories may suffice to explain the main 
features of outgoing gravitational radiation for head-on mergers.  
This would further imply that linear perturbations to exact black-hole 
solutions can capture the nonlinear aspects of head-on binary-black-hole
mergers accessible to observers far from the collision.

\end{abstract}

\pacs{04.25.Nx, 04.30.-w, 04.70.-s}

\maketitle

\section{Introduction}
\label{sec:introduction}

Even prior to the complete numerical-relativity simulations of 
black-hole-binary mergers \cite{Pretorius, Campanelli, Baker, Scheel},
black-hole collisions were thought to take place in three stages: 
inspiral (or infall), merger, and ringdown.  
During inspiral, the speed of the holes is sufficiently low and the 
separation of the bodies is large enough that the system behaves like
two separate particles that follow the post-Newtonian (PN) equations of motion.
Eventually, the black holes become sufficiently close that the dynamics given 
by the PN expansion significantly differ from those of full relativity.
This stage is the merger, during which the two black holes become
a single, highly distorted, black hole.  
The merger phase is brief; the strong deformations lose their energy to 
gravitational radiation, and the system settles down to a weakly perturbed, 
single black hole.
The ringdown phase describes these last small oscillations of the black hole.

Of the three stages of binary-black-hole coalescence, merger remains the most
inaccessible to analytical tools.
Nevertheless, full numerical relativity is not the only technique to have
success investigating merger.
Historically, most analytical investigations of the merger 
phase arise from trying to extend the validity of perturbative techniques, 
particularly black-hole-perturbation (BHP) and PN theories.
Those researchers working from a BHP approach try to push the approximation to
hold at earlier times, whereas those employing a PN method attempt to stretch
the technique to hold later into merger.
Alternatively, one can see if there are exact, nonlinear analytical models
whose dynamics can represent various aspects of black-hole-binary mergers.
Rezzolla, Macedo, and Jaramillo \cite{Rezzolla} recently took this latter
approach in their study of anti-kicks from black-hole mergers.
In their work, they showed that they could relate the curvature anisotropy 
on the past apparent horizon of a Robinson-Trautman spacetime to the kick 
velocity (computed from the Bondi momentum).
Through appropriate tuning of the initial data, they could recover kick 
velocities found in numerical-relativity simulations of unequal-mass,
insprialing black holes.
While this type of approach is interesting and has proved successful, our
work focuses on using perturbative approaches (and we will, therefore,
take a more comprehensive look at the prior use of perturbative methods to 
understand mergers).

From the BHP side, Price and Pullin \cite{Price}, initially, and many 
collaborators, subsequently, (see, e.g., \cite{Khanna}) developed the 
``close-limit approximation'' (CLA).
This technique begins with initial data containing two black holes that 
satisfy the vacuum Einstein equations and splits it into a piece representing 
the final, merged black hole and perturbations upon that black hole.
The exact form of the initial data varies in the CLA, but for
head-on collisions of black holes, it typically involves some variation of 
Misner \cite{Misner}, Lindquist \cite{Lindquist}, or Brill-Lindquist 
\cite{Brill} time-symmetric, analytic, wormhole-like solutions.
To investigate the late stages of an inspiral, the CLA usually begins from 
non-time-symmetric, but conformally flat, multiple black-hole initial data 
set forth by Bowen and York \cite{Bowen}.
Independent of the initial data, however, the CLA translates the original 
problem of merger into a calculation involving BHP theory.
The CLA does not allow for a very large separation of the black holes; as a 
result, only the very end of the merger is captured in this process.
Moreover, ``junk radiation'' appears in the waveform because the initial data 
do not describe the binary black-hole-merger spacetime in both the wave zone 
and the near zone.  
Unlike in full numerical-relativity simulations where the junk radiation
leaves the grid during the well understood inspiral phase, in the CLA the 
junk radiation appears during the merger stage.
This radiation (both from the absence of waves in the initial data and from 
errors in the near-zone physics), therefore, is difficult to disentangle 
from the physical waveform.

The Lazarus Project (see e.g.\ \cite{Baker2}) followed roughly the same 
approach as the CLA, but it used even more realistic black-hole-binary initial 
data for its CLA calculation; namely, its initial data came from a 
numerical-relativity simulation just prior to merger.
At the same time, however, because the initial data is now numerical, one
loses the analytical understanding of the properties of spacetime near
merger.
More recently, Sopuerta, Yunes, and Laguna \cite{Sopuerta} applied the CLA
in combination with PN flux formulae to obtain an estimate of the 
gravitational recoil from unequal-mass binaries (including binaries
with small eccentricity \cite{Sopuerta2}).
They proposed using more realistic initial data in the CLA, which 
Le Tiec and Blanchet \cite{LeTiec} ultimately carried out.
Le Tiec and Blanchet chose to use the 2PN metric (keeping only the first 
post-Minkowski terms) as initial data for the CLA, and they applied it with 
considerable success to inspirals of unequal-mass black-hole binaries in a 
paper with Will \cite{LeTiec2}.
Despite the improved initial data, this approach does not eliminate 
the problem of junk radiation discussed above.
It would be of interest to see if even more realistic initial data, such as that
of Johnson-McDaniel, Yunes, Tichy, and Owen \cite{Johnson} would lead to
improved results within the CLA.

From the PN side, Buonanno et al., \cite{Buonanno}, as well as 
Damour and Nagar \cite{Damour} take a different approach to understanding the 
physics of merger.
Using the Effective-One-Body (EOB) method \cite{Buonanno2}, 
they study the dynamics of the system until roughly the beginning
of the merger phase.
To obtain a complete waveform, they attach a ringdown waveform by smoothly 
fitting quasinormal modes to the EOB inspiral and plunge waveform.
When they calibrate the two free parameters of this model to 
numerical-relativity waveforms, the EOB results match numerical-relativity 
waveforms precisely.
In this method, they fit the PN dynamics and the ringdown at the light 
ring of the EOB particle motion; it is not immediately apparent, however,
what this feature tells about the nature of merger.

As a result, there remains a need to develop simple analytical models that
help reveal the behavior of spacetime during merger.
Toward this end, it is helpful to delve deeper into the question of what
exactly is the merger.
First, the inspiral-merger-ringdown classification is based on the validity 
of the PN expansion and that of BHP.
The inspiral, in other words, is just the set of times for which the 
PN expansion holds on the whole spatial domain (to a given level of accuracy).
Correspondingly, the ringdown is the times for which BHP
works everywhere throughout space.
Merger, in this picture, is just the gap between those times during which PN 
and BHP theories give accurate results.

In this paper, we propose that we can push each approximation technique 
beyond its current range of use, as long as we do not apply it to all of 
space at a given time.
We observe that at any time, there is a region outside a certain radius from 
the center of mass in which BHP applies.
While this seemingly runs contrary to the common notion that PN theory is the 
natural description of the weak-field region of a black-hole-binary spacetime,
a black-hole metric in the limit of radii much larger than the mass and binary
separation is identical to that of PN in the same region.
If the PN expansion applies to the remaining portion of the spacetime (within
the region where BHP holds), then BHP and PN theories could cover the entire 
physics of black-hole-binary coalescence.
While it is somewhat unreasonable to suppose that PN theory truly applies
to the strong-field region of a binary-black-hole spacetime, revisiting
Price's treatment of non-spherical stellar collapse suggests
that this may not be essential.

In Price's 1972 paper \cite{Price2}, he addresses, among other issues, why
aspherical portions of stellar collapse quickly disappear when, in fact,
one could plausibly argue the contrary.
Namely, if any non-spherical perturbations asymptote to the horizon (from the
perspective of an observer far away), they would remain visible to this
observer indefinitely.
Price realizes, however, that the exterior of a collapsing star is just the
Schwarzschild solution (due to Birkhoff's theorem, up to small perturbations),
and perturbations on the Schwarzschild spacetime evolve via a radial wave
equation in an effective potential.
Moreover, he notes that when the surface of the star passes through the 
curvature effective potential of the Schwarzschild spacetime, the 
gravitational perturbations induced by the star redshift.
Finally, because the effective potential reflects low-frequency perturbations,
the spacetime distortions produced by the stellar interior become less 
important, and observers outside the star ultimately see it settle into 
a spherical black hole in a finite amount of time.
Most importantly, this argument does not depend strongly upon the physics of 
the stellar interior; as long as there is gravitational collapse to a black 
hole, this idea holds.

\begin{figure}
\includegraphics[width=0.45\textwidth]{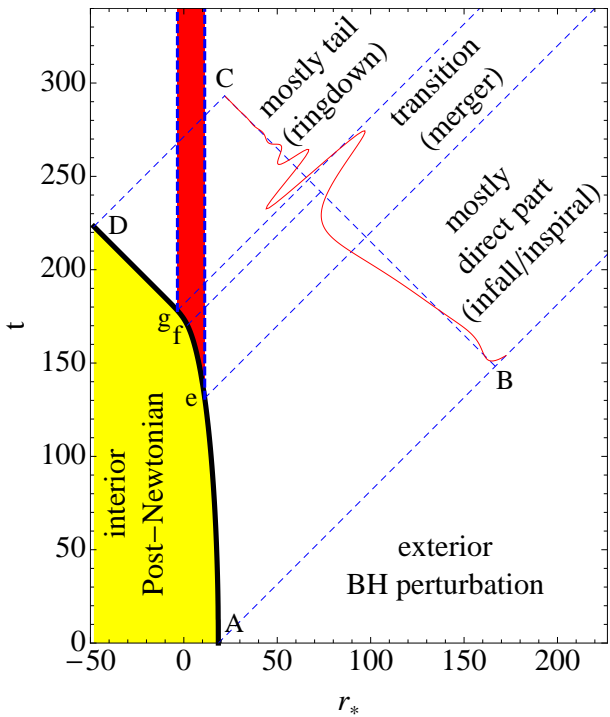}
\caption{(Color online) This figure depicts a spacetime diagram of a 
black-hole collision, modeled after Price's description of stellar collapse.
We choose the trajectory of the two holes as a way to separate the spacetime
into an interior and an exterior region.
The exterior region is a perturbed, black-hole spacetime, whereas the
interior is that of a post-Newtonian (PN) black-hole-binary system
[shaded in yellow (light gray)].
The red (dark gray) region of the diagram shows the place at which the
effective potential of the black hole is significantly greater than zero.
This formalism allows us to divide the waveform into three sections:
inspiral (or infall), which extends from the beginning of the binary's
evolution until when the $l=2$ effective potential of the exterior BHP 
spacetime starts to be exposed; merger, which extends from the end of inspiral 
to when the majority of the exterior potential is revealed; and ringdown, 
which represents the remainder.
We overlay the even-parity, ($l=2$,$m=0$) mode of the waveform.
\label{fig1}}
\end{figure}

In this paper, we adopt this idea, but we replace the stellar physics of the
interior with a PN, black-hole-binary spacetime (see Fig.~\ref{fig1}).  
While in Price's case, the division of spacetime into two regions comes
naturally from tracking the regions of space containing the star and vacuum,
in our case the split is somewhat more arbitrary; one simply needs to find a 
region in which both PN and BHP theories hold, to some
level of accuracy.
How we choose the boundary between the two regions and the quantities that
we evolve are topics that will be discussed in greater detail in
Sec.~\ref{sec:qualitative}.

To test the above idea in this paper, we study a head-on collision of two 
black holes with transverse, anti-aligned spins and compare the waveforms and 
energy-momentum flux obtained from our approximation method with the 
equivalent quantities from full numerical simulations.
Specifically, we organize this paper as follows.
In Sec.~\ref{sec:qualitative}, we give a more detailed motivation for our 
model, and we then present the mathematical procedure we use in our method,
for an equal-mass, head-on collision of black holes.
In Sec.~\ref{sec:implementation}, we present an explicit calculation for the 
head-on collision of two black holes with transverse, anti-aligned spins, and 
we compare waveforms, radiated energy and radiated linear momentum, from our 
model with the equivalent quantities from a full numerical-relativity 
simulation.
In Sec.~\ref{sec:discussion} we discuss how our method can help interpret 
the waveform during merger, and finally, in Sec.~\ref{sec:conclusion}, 
we conclude.
We will use geometrical units ($G=c=1$) and Einstein summation convention
throughout this paper.

\section{A Detailed Description of the Method}
\label{sec:qualitative}

\subsection{Further Motivation}

Before going into the details of our procedure, it is worth spending some time
discussing why our specific implementation of PN and BHP theories will help
avoid some of the difficulties that arose in other methods in the introduction
and noting the limitations and assumptions that underlie our approach.

It is certainly hard to argue that existing orders of PN (up to $v^6$ in the
metric, for near-zone dynamics \cite{Blanchet}) and BHP
(up to second order for Schwarzschild, see \cite{Garat} for a gauge-invariant
formulation) theories are accurate in the whole space, simultaneously.
Nevertheless, it is plausible to argue that these approximation techniques
cover different spatial regions at different times in a way such that each
theory is either valid to a reasonable level of accuracy or occupying a
portion of spacetime that will not influence physical observables where
it fails.
Using an approach of this type, we aim to get the most out of the 
approximation methods.

Specifically, we find that the following procedure gives good agreement with 
the waveform of a numerical-relativity simulation presented in Sec.\ 
\ref{sec:implementation}.
First, we have the reduced mass of the binary follow a plunging geodesic in
the Schwarzschild spacetime.
Then, we divide this trajectory in half to make a coordinate radius (and thus
a coordinate sphere) that passes through the centers of the black holes.
The set of all the coordinate spheres defines a time-like surface in spacetime.
Finally, we apply PN theory within this time-like surface and BHP on the
exterior.
The two theories must agree on this time-like surface, which we will 
subsequently call the shell.

Matching PN and BHP theories on this shell has certain advantages.
Because BHP theory relies upon a multipole expansion, this makes it necessary 
to treat the PN interior in terms of multipoles of the potentials.
For one, this is useful, because physical observables like the radiated 
energy and momentum very often do not need many multipoles to find
accurate results.
(In fact, in our example in Sec.\ \ref{sec:implementation}, we see that
the quadrupole perturbations alone suffice.)
Second, a multipole expansion may also be helpful for the convergence of
the approach.
For two point particles, for example, each multipole component of the 
Newtonian potential $U_N^{(l)}$ at the location of the particles satisfies 
$U_N^{(l)}\lesssim M/R$, where $M$ is the mass of the binary and $R$ is the 
distance from the center of mass.
This is small for much of the infall, when $R \gg M$, and even when the binary
reaches what will be the peak of the effective potential of the merged
black hole, $U_N^{(l)} \sim 1/3$.

Because the effective potential of the final black hole tends to mask 
perturbations within (as they are redshifted and cannot escape), our hope 
(supported by the example in Sec.\ \ref{sec:implementation}) is that PN theory 
is still reasonably accurate at the peak of the potential.
Then, in our approach, the PN error will be hidden by this potential and, 
along with BHP in the exterior, it will suffice to explain the physics outside 
the black hole (in particular the gravitational waveform and the energy and
momentum flux).
Of course whether such a mechanism exists is not easily deduced analytically 
from first principles; only by comparing our results with those from full 
numerical relativity will we test this premise (which we do in Sec.\
\ref{sec:comparison}).

If this holds, our procedure has several advantages.
For one, the matching works well at larger separations (which diminishes the
influence of junk radiation).
More importantly, though, as the matching shell moves from large separations
at early times to the vicinity of the horizon at late times, the
spacetime smoothly transitions from predominantly PN to essentially BHP.
One can see this most clearly in the way that the effective potential is 
revealed in Fig.\ \ref{fig1}.
This leads to a waveform that smoothly transitions between an inspiral at
early times to a ringdown at late times.
Whether we correctly capture the merger phase is most easily confirmed by
comparing with results from numerical relativity.
Finally, our method can give a way to interpret the waveform during the 
different stages of a binary-black-hole coalescence by relating parts of the 
waveform with the retarded position of the matching shell at that point 
(for further detail, see Sec.\ \ref{sec:discussion}).

There is also reason to suspect that the domain of BHP may extend to 
relatively early times in the merger, outside the binary.
A paper by Racine, Buonanno and Kidder \cite{Racine} gives evidence
in favor of this idea.
In particular, they find, when studying the ``super-kick'' configuration 
(an equal-mass, circular binary of black holes with transverse, anti-parallel
spins), that the higher-order spin-orbit contributions to linear-momentum flux 
dominate over the leading-order terms.
These terms include both direct higher PN terms and tail terms, where the tail
refers to gravitational radiation that is scattered off of spacetime curvature 
and propagates within the light cone (as opposed to the direct piece that 
propagates on the light cone).
This suggests that even early on, the background curvature plays an important
role in generating the kick.
Whether a Schwarzschild black hole properly represents this curvature is 
another idea that is difficult to argue for directly, but can be confirmed by 
comparison with the results of numerical relativity.

In its current implementation, the feature of our approach that is most
arbitrary is setting the boundary between the PN interior and the BHP
exterior.
Nevertheless, our procedure works for head-on collisions, and in future work,
we will develop a more systematic way of treating the boundary.

\subsection{Details of the Implementation}
\label{sec:details}

The procedure that we follow can be broken down, more or less, into five
steps: 
(i) we describe relevant physics in the PN interior;
(ii) we match the metric of the PN interior to the BHP metric through a
boundary; 
(iii) we explicitly construct the boundary between the PN and BHP spacetimes;
(iv) we evolve the metric perturbations in the exterior Schwarzschild 
spacetime;
(v) we extract the waveforms and compute the radiated energy and momentum.
We shall devote a subsection to each of these topics below.

Before we do so, however, it will be helpful to clarify our notation regarding
the different coordinates we use for the two metrics and the matching shell.
In the PN coordinate system, we use Minkowski coordinates $(T,X,Y,Z)$ and 
spherical-polar coordinates $(R,\Theta,\Phi)$ within spatial slices.
We will only consider linear perturbations to Minkowski in the harmonic gauge.
For our BHP, we employ Schwarzschild coordinates $(t,r,\theta,\varphi)$, and 
similarly, we only examine linear perturbations to the Schwarzschild
spacetime. 
As we will show in Sec.\ \ref{sec:shell}, we can match these descriptions 
when $R,r \gg M$.
This procedure is accurate up to terms of order $(M/R)^2$ in the monopole 
part of the metric and $M/R$ in the higher-multipole portions, assuming that
we relate the two coordinate systems by
\begin{equation}
T=t\,,\;\Theta=\theta\,,\;\Phi=\varphi\,,R=r-M \, .
\end{equation}
The identification above allows us to label every point in spacetime by two 
sets of coordinates, $(t,r,\theta,\varphi)$ and  $(t,R,\theta,\varphi)$, 
where $R=r-M$. 
Because our program relies upon applying PN theory in an interior region
and BHP on the exterior, it is therefore natural to talk about a coordinate 
shell at which we switch between PN and BHP descriptions of the space-time.
On this shell, we can use either the Minkowski or Schwarzschild coordinates.

\begin{figure}[htb]
\includegraphics[width=0.45\textwidth]{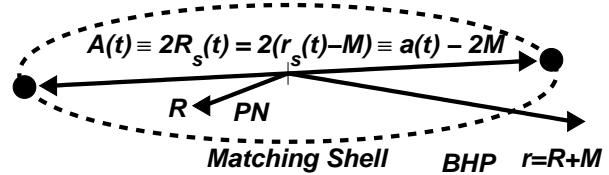}
\caption{This figure shows, at a given moment in time, the Schwarzschild and 
PN radial coordinates, the binary separation, and the position of the shell 
where we match the two theories.}
\label{r_coords}
\end{figure}

In keeping with the notation above, we shall denote the separation of the
binary by $A(t) = 2R(t)$ in PN coordinates and $a(t) = 2r(t)$ in 
Schwarzschild coordinates.
Finally, we will denote the radial coordinate on the boundary by adding
a subscript $s$ to the PN or Schwarzschild coordinate radii [e.g., 
$R_s(t) = A(t)/2$ or $r_s(t) = a(t)/2$].
We summarize the two coordinate systems and how they match, in Table 
\ref{notation_table} and Fig.\ \ref{r_coords}.

\begin{table*}[htb] 
\begin{tabular}{cccc}
\hline
\hline
& PN Spacetime & Matching Shell & Perturbed Schwarzschild Spacetime\\
\hline
Coordinates & $(t,R,\theta,\varphi)$ & $(t,R_s(t),\theta,\varphi)$ or
$(t,r_s(t),\theta,\varphi)$ & $(t,r,\theta,\varphi)$, $r = R+M$\\
Binary Separation & $A(t)$ & $A(t)$ or $a(t)$ & $a(t)$ \\
Matching Radius & $\quad R(t) = A(t)/2 \quad$ & $R_s(t) = a(t)/2 - M$ or
$r_s(t) = A(t)/2 + M$ & $r(t) = a(t)/2$ \\
\hline
\hline
\end{tabular}
\caption{This table summarizes our notation for the coordinates, the binary
separation, and the matching radius that we use for the PN spacetime, the
BHP spacetime, and the matching region between the two.}
\label{notation_table}
\end{table*}

While we introduce a new method of matching PN and BHP theories, the idea 
of combining PN and BHP approximations is not new.  
In fact, it is at the core of the Effective-One-Body formalism of Buonanno 
and Damour \cite{Buonanno2}.  
In the EOB description, however, they match the point-particle Hamiltonians
of PN and BHP theories, rather than joining the spacetime geometry.
It would be interesting, as a future study, to see whether one can combine
our procedure with that of the EOB to produce a geometrical EOB approach.  
Le Tiec and Blanchet \cite{LeTiec}, on the other hand performed a more 
accurate matching between PN and BHP in their close-limit calculation with 
2PN initial data, but their matching only takes place on a single spatial
slice of initial data.
It would also be interesting to extend their higher-order approach to our
procedure as well.

\subsubsection{The PN Interior Solution}

For our method, we will need to describe the metric of the PN spacetime
in the interior, which we do at leading Newtonian order:
\begin{eqnarray}
\nonumber
ds^2_{\rm PN} &=& -(1-2U_N)dt^2 - 8w_i dt dX^i \\
& +& (1+2U_N)\delta_{ij}dX^i dX^j \, .
\end{eqnarray}
Here  $U_N$ is the Newtonian potential and $w_i$ is the gravitomagnetic
potential, and the index $i$ runs over $X$, $Y$, and $Z$.
Our notation follows \cite{Kaplan} [the above takes the results of 
Eq.\ (2.1) of that paper].
We then expand the Newtonian potential, $U_N$, into multipoles, keeping only
the lowest multipoles necessary to complete the calculation.
In this paper the monopole and quadrupole pieces suffice 
(the dipole piece can always be made to vanish),
\begin{equation}
U_N \approx  U_N^{(0)} + U_N^{(2)} \, .
\end{equation}
The quadrupole piece can be expressed as a term without angular dependence
times a spherical harmonic, as is done below,
\begin{equation}
U_N^{(2)} = U_N^{2,m} Y_{2,m}(\theta,\varphi) \, .
\end{equation}
We shall follow the same procedure with the gravitomagnetic potential,
although here we will, temporarily, keep the dipole term,
\begin{equation}
w_i \approx w_i^{(1)} + w_i^{(2)} \, .
\end{equation}
We will be able to remove the dipole term through gauge transformations,
but this discussion is much simpler on a case-by-case basis.
When we write the gravitomagnetic potential, ${\bf w}$, in spherical polar
coordinates, we will be able to remove the radial component through a
gauge transformation.
We will, therefore, consider just the $\theta$ and $\varphi$ components
of ${\bf w}$, and when writing it in index notation, we will denote
them with Latin letters from the beginning of the alphabet, 
(e.g.\ $a,b = \theta,\varphi$).
We can then expand the components $w_a^{(2)}$ in terms of two
vector spherical harmonics,

\begin{equation}
w_a^{(2)} = w^{2,m}_{(\rm e)} \nabla_a Y_{2,m}(\theta,\varphi) +
w^{2,m}_{(\rm o)} \epsilon_a\,^b \nabla_b Y_{2,m}(\theta,\varphi) \, .
\end{equation}

Here $\nabla_a$ is the covariant derivative on a 2-sphere, and 
$\epsilon_a\,^b$ is the Levi-Civita tensor (with nonzero components
$\epsilon_\theta\,^\varphi = 1/\sin\theta$ and 
$\epsilon_\varphi\,^\theta = -\sin\theta$).
A convenient abbreviation for the two spherical harmonics above is
\begin{equation}
w_a^{(2)} = w^{2,m}_{(\rm e)} Y^{2,m}_a + w^{2,m}_{(\rm o)} X^{2,m}_a \, .
\end{equation}
The two terms are denoted by $(\rm e)$ and $(\rm o)$ as a short hand for even 
and odd, because they transform as $(-1)^l$ and $(-1)^{l+1}$ under parity
transformations, respectively.
The odd- and even-parity vector spherical harmonics are given explicitly by
\begin{equation}
X_\theta^{l,m} =-\frac{1}{\sin\theta}\frac{\partial Y^{lm} }{\partial\varphi }
\,,\quad
X_\varphi^{l,m} ={\sin\theta}\frac{\partial Y^{lm}}{\partial\theta }
\end{equation}
and
\begin{equation}
Y_\theta^{l,m} =\frac{\partial Y^{lm} }{\partial\theta}
\,,\quad
Y_\varphi^{l,m} =\frac{\partial Y^{lm}}{\partial\varphi} \, ,
\end{equation}
respectively.
These are the only parts of the PN metric that will be necessary for our 
approach.

\subsubsection{Matching to Perturbed Schwarzschild}

We then note that the Schwarzschild metric takes the form
\begin{equation}
ds^2 = -\left(1-\frac{2M}r\right) dt^2 + \left(1-\frac{2M}r
\right)^{-1} dr^2 + r^2 d^2\Omega \, ,
\end{equation}
where the last piece is the metric of a 2-sphere.
We will use $r$ without any subscript to denote the Schwarzschild radial
coordinate.
When $M \ll r$ the Schwarzschild metric takes the form
\begin{equation}
ds^2 \approx -\left(1-\frac{2M}r\right) dt^2 + \left(1+\frac{2M}r
\right) dr^2 + r^2 d^2\Omega \, .
\end{equation}
By making the coordinate transformation $R = r - M$, and identifying
$M/R$ with the monopole piece of the Newtonian potential, $U_N^{(0)}$, 
then one can find that the Schwarzschild metric takes 
the form of the Newtonian metric in spherical coordinates,
\begin{eqnarray}
\nonumber
ds^2 &\approx& -(1-2U_N^{(0)}) dt^2 
+ (1+2U_N^{(0)}) \\
&&\times (dR^2 + R^2 d^2\Omega) \, .
\end{eqnarray}

This similarity between the PN and Schwarzschild metrics suggests a way
to match the two at the boundary.
We will assume that the monopole piece of the Newtonian potential 
becomes the $M/r$ term in the Schwarzschild metric.
For the remaining pieces of the Newtonian metric (namely $U_N^{(2)}$ and 
$w_i^{(2)}$) we will translate them directly into the Schwarzschild metric
after performing any needed gauge transformations to make such a direct
match reasonable.

The original works on perturbations of the Schwarzschild spacetime are
those of Regge and Wheeler \cite{Regge} for the odd-parity perturbations
and Zerilli \cite{Zerilli} for the even-parity perturbations.
Moncrief \cite{Moncrief} then used a variational principle to show that 
one can derive quantities from the metric perturbations of Regge, Wheeler
and Zerilli that satisfy a well-posed, initial-value problem in any 
coordinates that deviate from Schwarzschild at linear order in perturbation
theory.
These quantities are related to the gravitational waves at infinity, and
they satisfy a one-dimensional wave equation in a potential.
We follow Moncrief's approach in computing these so-called gauge-invariant 
metric perturbation functions, but for our notation, we use that of
a recent review article by Ruiz et al.\ \cite{Ruiz}.

Both the even-parity [transform as $(-1)^l$ under parity] and odd-parity
[transform as $(-1)^{l+1}$ under parity] perturbations are not very difficult
to find.
By writing the PN metric in spherical-polar coordinates,
\begin{eqnarray}
\nonumber
ds^2 &=& -(1-2U_N^{(0)}-2U_N^{(2)})dt^2 + (1+2U_N^{(0)}+2U_N^{(2)})\\
&&\times(dR^2+R^2d^2\Omega) -8w_b^{(2)}dtdx^b \, ,
\end{eqnarray}
where $dx^b = d\theta, d\varphi$, one can see that the even-parity 
perturbations are nearly diagonal in the metric.
In fact, at leading Newtonian order, it is exactly diagonal, because the
non-diagonal term coming from $w^{2,m}_{(\rm e)}$ arises at a higher
PN order.
We will show this explicitly in Sec.\ \ref{sec:implementation}.
For this reason, we only consider the diagonal metric components in the
discussion below.

The even-parity metric perturbations in Schwarzschild are often denoted
\begin{eqnarray}
(h_{tt}^{l,m})_{(\rm e)} = H_{tt}^{l,m} Y^{l,m} \, , &\,&
(h_{rr}^{l,m})_{(\rm e)} = H_{rr}^{l,m} Y^{l,m} \, ,\\
\nonumber
(h_{\theta\theta}^{l,m})_{(\rm e)} =  r^2 K^{l,m} Y^{l,m} \, , &\,&
(h_{\varphi\varphi}^{l,m})_{(\rm e)} =  r^2 \sin^2\theta K^{l,m} Y^{l,m} \, ,
\end{eqnarray}
a specialization of Eqs.\ (57)-(59) of Ruiz et al.
Thus, by matching the two metrics, one can see that
\begin{equation}
H_{tt}^{2,m} = H_{rr}^{2,m} = K^{2,m} = 2U_N^{2,m} \, .
\label{even_perturbs}
\end{equation}
The odd-parity term is somewhat simpler, because there is only
one metric perturbation in Schwarzschild to match at leading order,
\begin{equation}
(h_{t\theta}^{l,m})_{(\rm o)} = h_t^{l,m} X_{\theta}^{l,m} \, , \qquad
(h_{t\varphi}^{l,m})_{(\rm o)} = h_t^{l,m} X_{\varphi}^{l,m} \, ,
\end{equation}
Eq. (61) of Ruiz et al. 
From this, one can find that 
\begin{equation}
h_t^{2,m} =-4w^{2,m}_{(\rm o)} \, .
\label{odd_perturbs}
\end{equation}
The matching procedure thus gives a way to produce perturbations in
the Schwarzschild spacetime.

\subsubsection{The Boundary Shell}
\label{sec:shell}

We now must find a boundary region where one can match a PN metric
expanded in multipoles with a perturbed Schwarzschild metric.
For head-on collisions, we find that the boundary can be a spherical shell 
whose radius varies in time as the binary evolves.
We can motivate where this boundary should be with a few simple arguments,
but the true test of the matching idea will come from comparisons with
exact waveforms from numerical relativity.

We know that at early times and for larger separations of the black holes,
the PN spacetime is valid around the two holes; thus it is not unreasonable to
suppose that the shell should have a radius equal to half the binary
separation.
Later in the evolution, BHP will be valid everywhere, so the shell should
asymptote to the horizon of the merged hole (as seen by outside observers).
The trajectory of the shell should be smooth throughout the entire
process, as well.
Finally, the boundary should mimic the reduced-mass motion of the system,
which physically generates the gravitational waves.

A simple way to achieve this quantitatively is instead of having the motion
of the reduced mass follow the PN equations of motion, we impose that it
undergoes plunging geodesic motion in the Schwarzschild spacetime.
Given that the exterior spacetime is a perturbed Schwarzschild, and that we
are matching the two approximations on a shell that passes through the centers
of the two black holes, it is just as reasonable to use a trajectory in the
Schwarzschild spacetime.
Moreover, at large separations, the motion of the reduced mass of the system
in both Schwarzschild and PN are quite similar; we primarily choose the
geodesic in Schwarzschild for its behavior at late times.
For completeness, we write down the differential equation we use to find the
motion of a radial geodesic in Schwarzschild.
Since we think of the black holes as point particles residing in the PN 
coordinate system, we write the evolution of the binary separation $A(t)$, 
measured in the PN coordinates,
\begin{eqnarray}
\nonumber
\frac{dA(t)}{d\tau} &=& -\sqrt{B^2 - (1-2M/A(t))}  \, ,\\
\frac{dt}{d\tau} &=& B\left(1-\frac{2M}{A(t)}\right)^{-1} \, ,
\end{eqnarray}
where $B$ is a positive constant ($B=1-2M/A(0)$ for a head-on plunge from rest).
This expression can be found in many sources; see, for example \cite{Stephani}.
The coordinate radius of the shell in the PN spacetime is just half the 
distance $A(t)$,
\begin{equation}
R_s(t) = \frac 12 A(t) \, .
\end{equation}

Since the PN and Schwarzschild radii are related by $R = r - M$, one can find
that in the Schwarzschild coordinates, the position of the shell is given by:
\begin{equation}
r_s(t) = \frac{1}{2} A(t) + M \, .
\end{equation}
In the Schwarzschild coordinates, since $A(t)$ goes to $2M$ at late times,
the radius of the shell asymptotes to the horizon.  
More specifically, let us first define $\delta r_s(t) = r_s(t) - 2M$ and then
note that at late times, the trajectory approaches the speed of light as
it falls toward the horizon.
In terms of the tortoise coordinate, $r_* = r + 2M\log[r/(2M)-1]$, this 
occurs when $v = t + r_* = {\rm const.}$.
Writing this with respect to the variable $\delta r_s(t)$, we find 
\begin{equation}
-\frac t{2M} \sim \frac{\delta r_s(t)}{2M} + \log \left(
\frac{\delta r_s(t)}{2M} \right) \, .
\end{equation}
We neglected the constant value of $v$ because doing so has no affect on 
finding the scaling of $r_s(t)$ at late times $t$.
The equation above has a solution in terms of the Lambert W function, $W(x)$,
given by
\begin{equation}
\delta r_s(t) \sim W(e^{-t/(2M)} ) \, .
\end{equation}
Because we are interested in the behavior at large $t$, $e^{-t/(2M)}$ is 
small, and we can use the leading-order term in the Taylor series for the
Lambert $W$ function, $W(x) \sim x + O(x^2)$.
We find that
\begin{equation}
\delta r_s(t) \sim e^{-t/(2M)} \, .
\end{equation}

\begin{figure}[htb]
\includegraphics[width=0.45\textwidth]{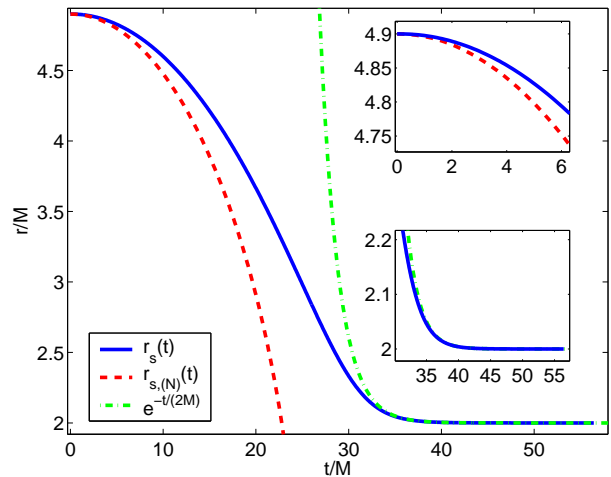}
\caption{(Color online) This figure shows the trajectory of the boundary 
shell as the solid blue (black) curve labeled by $r_s(t)$.  
The other two curves show the early- and late-time behavior of the shell.
The red (gray) dashed curve labeled by $r_{s,(N)}(t)$ shows the trajectory
of a shell that follows the Newtonian equations of motion for a plunge from
rest.
The green (light gray) dashed and dotted curve [denoted by $e^{-t/(2M)}$] 
shows the exponential convergence to the horizon at the rate expected
in a Schwarzschild spacetime.
The upper inset shows how the shell agrees with a Newtonian plunge from
rest at early times, and the lower inset shows how the shell converges
exponentially to the horizon at the expected rate.}
\label{trajectories}
\end{figure}

Thus, matching the spacetimes at half the PN separation of the binary and 
having the separation of the binary evolve via a Schwarzschild geodesic in
the PN coordinate system makes the shell track the PN reduced-mass motion
at early times, but still head to the horizon at late times.
We illustrate these different behaviors by plotting the full trajectory of
the shell $r_s(t)$, the trajectory of a shell that follows a plunging
Newtonian orbit (which we denote by $r_{s,(N)}(t)$ and which represents the 
behavior of the shell at early times), and $e^{-t/(2M)}$ (the late-time
behavior of the shell) in Fig.\ \ref{trajectories}.
We choose the initial values of the trajectory to conform with those of the 
numerical simulation with which we compare in Sec.\ \ref{sec:comparison}.
The upper and lower insets show how the shell's trajectory converges to
the Newtonian behavior (at early times) and the expected exponential decay
at late times.

While choosing the trajectory of the shell might have a slightly ad hoc feel,
in future work we will develop a framework that determines the shell motion
consistently through radiation reaction.

\subsubsection{Black-Hole Perturbations}

One can then take the odd- and even-parity metric perturbations from the 
second subsection and transform them into two quantities, the Regge-Wheeler
and Zerilli functions, respectively, that each satisfy a simple wave equation.
We first treat the even-parity perturbations.
Eqs.\ (63)-(65) of Ruiz et al.\ show how to take metric perturbations and
transform them into the even-parity, gauge-invariant Zerilli function.
Substituting our Eq.\ (\ref{even_perturbs}) into those three of Ruiz, we find
\begin{eqnarray}
\nonumber
\Psi^{2,m}_{(\rm e)} &=& \frac{2r}{3} \left\{U_N^{2,m} + \frac{r-2M}{2r+3M}
\right.\\
&&\times\left.\left[\left(1-\frac{2M}{r}\right) U_N^{2,m} - 
r\partial_r U_N^{2,m}\right]\right\} \, ,
\label{z_function_gen}
\end{eqnarray}
where $\partial_r$ is just the radial derivative with respect to the 
Schwarzschild radial variable.
The odd-parity perturbations come directly from applying our Eq.\ 
(\ref{odd_perturbs}) to Eq.\ (67) of Ruiz et al.
This gives that
\begin{equation}
\Psi^{2,m}_{(\rm o)} = 2r\left(\partial_r w^{2,m}_{(\rm o)} - \frac{2}{r}
w^{2,m}_{(\rm o)}\right) \, .
\label{rw_function_gen}
\end{equation}

The odd- and even-parity perturbations then evolve according to the 
Regge-Wheeler and Zerilli equations respectively,
\begin{equation}
(\partial_t^2 - \partial_{r_*}^2 + V_{(\rm{e,o})}^l(r))\Psi_{(\rm{e,o})}^{l,m} 
= 0 \, ,
\end{equation}
where $r_* =r+2M\log[r/(2M)-1]$ is the tortoise coordinate.
The potentials can be expressed most concisely via the expression
\begin{equation}
V_{(\rm{e,o})}^l(r) = \left(1-\frac{2M}r\right) \left(\frac{\lambda}{r^2} -
\frac{6M}{r^3}U^l_{(\rm{e,o})}(r)\right) \, ,
\end{equation}
where $\lambda = l(l+1)$ and 
\begin{equation}
U^l_{(\rm o)}(r) = 1, \quad U^l_{(\rm e)}(r) = \frac{\Lambda(\Lambda+2)r^2+
3M(r-M)}{(\Lambda r+3M)^2} \, .
\end{equation}
Here $\Lambda = (l-1)(l+2)/2 = \lambda/2 - 1$.
These expressions follow Eqs.\ (5.3)-(5.6) of \cite{LeTiec}.

In our procedure, we find it easiest to evolve the Regge-Wheeler-Zerilli
equations using a characteristic method.
To do so, we define $u=t-r_*$ and $v=t+r_*$ and see that the evolution
equation becomes
\begin{equation}
\frac{\partial^2 \Psi^{l,m}_{\rm (e,o)}}{\partial u\partial v} 
+\frac{V_{\rm (e,o)}^l\Psi^{l,m}_{\rm (e,o)}}{4} = 0 \, .
\label{wave_eq}
\end{equation}
We will now discuss how we evolve our Regge-Wheeler-Zerilli functions
with the aid of Fig.\ \ref{fig1}.

We must provide data in two places, the matching shell (in Fig.\ \ref{fig1} 
it is the lower-left curve labeled by the points $AefgD$) and the initial 
outgoing characteristic (the line labeled by $AB$ on the lower right).
Once we do this, however, we can determine the Regge-Wheeler-Zerilli functions
within the quadrilateral (with the one curved side) $ABCD$.
We already discussed how we determine the shell in Sec.\ \ref{sec:shell},
and the data we provide along that curve are just the Regge-Wheeler
[Eq.\ (\ref{rw_function_gen})] or Zerilli [Eq.\ (\ref{z_function_gen})]
functions restricted to that curve.
The data we must provide along $AB$ are less well determined.
If our computational grid extended to past null infinity, then we could
impose a no ingoing wave condition.
At finite times, we can still impose this boundary condition, but it 
leads to a small spurious pulse of gravitational radiation at the beginning 
of our evolution.
To limit the effects of this, we keep the shell at rest until the junk 
radiation dissipates, and then we begin our evolution.
At this point, the data along the line $AB$ more closely represent those
of a binary about to begin its plunge.

With these data, we can then evolve the Regge-Wheeler-Zerilli equations 
numerically, using the second-order-accurate, characteristic method described 
by Gundlach et al.\ in \cite{Gundlach}.
The essence of this method is that one can use the data at a point plus
those at one step ahead in $u$ and $v$, respectively, to get the next
point advanced by a step ahead in both $u$ and $v$.
Explicitly, if one defines $\Psi_N = \Psi_{\rm (e,o)}^{l,m}(u+\Delta u,
v+\Delta v)$, $\Psi_W = \Psi_{\rm (e,o)}^{l,m}(u+\Delta u,v)$, 
$\Psi_E = \Psi_{\rm (e,o)}^{l,m}(u,v+\Delta v)$, and 
$\Psi_S = \Psi_{\rm (e,o)}^{l,m}(u,v)$, then one has that
\begin{eqnarray}
\nonumber
\Psi_N &=& \Psi_E + \Psi_W - \Psi_S - \frac{\Delta u\Delta v}{8} 
V^l_{\rm (e,o)}(r_c) (\Psi_E + \Psi_W)\\
&& + O(\Delta u^2 \Delta v, \Delta u \Delta v^2) \, ,
\end{eqnarray}
where $r_c$ is the value of $r$ at the center of the discretized grid.
Because our boundary data do not always lie on one of the grid points in 
the $(u,v)$-plane, we must interpolate the bottom point 
$\Psi_{\rm (e,o)}^{l,m}(u,v')$ to fall at the same value of $v$ as the next
boundary point at $\Psi_{\rm (e,o)}^{l,m}(u+\Delta u,v)$.
As long as we do this interpolation with a quadratic or a polynomial of 
higher degree, it does not seem to influence the second-order convergence of 
the method.
Finally, we can extract the Regge-Wheeler-Zerilli functions from the line 
$BC$ in Fig.\ \ref{fig1} as they propagate toward future null infinity.

\subsubsection{Waveforms and Radiated Energy-Momentum}

As we mentioned at the end of the previous section, it is easy to find
the Regge-Wheeler-Zerilli functions from the exterior of our computational
grid; this is useful, because these functions are directly related to the
gravitational waveform $h$, asymptotically. 
For radii much larger than the reduced gravitational wavelength,
$r\gg \lambda_{\rm GW}/(2\pi)$, one has that
\begin{eqnarray}
&&h_+ - i h_\times \nonumber \\
& =&\frac{1}{2r}\sum_{l,m} \sqrt{\frac{(l+2)!}{(l-2)!}}\left[\Psi^{l,m}_{(\rm e)}
+i\,\Psi^{lm}_{(\rm o)}\right]{}_{-2}Y_{lm} \, ,
\label{h_plus}
\end{eqnarray}
where ${}_{-2}Y_{lm}$ is a spin-weighted spherical harmonic.
The above comes from Eq.\ (84) of Ruiz et al., which also contains a
discussion about the spin-weighted harmonics in an appendix.
One can then substitute this into the usual expressions for the energy and
momentum radiated by gravitational waves,
\begin{eqnarray}
\frac{dE}{dt} &=& \lim_{r\rightarrow\infty} \frac{r^2}{16\pi} 
\oint |\dot h_+ - i\dot h_{\times}|^2 d\Omega \, ,\\
\frac{dP_i}{dt} &=& \lim_{r\rightarrow\infty} \frac{r^2}{16\pi} 
\oint n_i |\dot h_+ - i\dot h_{\times}|^2 d\Omega \, ,
\end{eqnarray}
(where $n_i$ is a unit vector, a dot represents a time derivative, 
 and $d\Omega$ is the volume element on a 2-sphere).
A lengthy, but straight-forward calculation done by
Ruiz et al.\ shows that
\begin{equation}
\frac{dE}{dt} = \frac{1}{64\pi} \sum_{l.m} \frac{(l+2)!}{(l-2)!} \left(
|\dot\Psi^{l,m}_{(\rm e)}|^2 + |\dot\Psi^{l,m}_{(\rm o)}|^2\right) 
\end{equation}
[their Eq.\ (91)].
For the components of the momentum we are interested in (in the $xy$-plane)
combining their Eqs.\ (86)-(88) and (93) and using their definition
$P_+ = P_x+iP_y$ gives
\begin{eqnarray}
\nonumber
\frac{dP_+}{dt} &=& -\frac{1}{16\pi} \sum_{l.m} \frac{(l+2)!}{(l-2)!} 
\left[ia_{l,m}\dot\Psi^{l,m}_{(\rm e)} \dot{\bar\Psi}^{l,m+1}_{(\rm o)} 
\right.\\ 
\nonumber
&&+b_{l+1,m+1}\left( \dot\Psi^{l,m}_{(\rm e)} 
\dot{\bar\Psi}^{l+1,m+1}_{(\rm e)}\right.\\
&&\left.\left.+ \dot\Psi^{l,m}_{(\rm o)} \dot{\bar\Psi}^{l+1,m+1}_{(\rm o)}
\right) \right] \, .
\end{eqnarray}
The bar denotes complex conjugation.
The coefficients $a_{l.m}$ and $b_{l,m}$ are given by their Eqs.\ (41) 
and (42), which we reproduce here
\begin{eqnarray}
a_{l.m} & = & \frac{\sqrt{(l-m)(l+m+1)}}{l(l+1)}\\
b_{l,m} & = & \frac 1{2l} \sqrt{\frac{(l-2)(l+2)(l+m)(l+m-1)}{(2l-1)(2l+1)}}
\end{eqnarray}
With the framework now in place, we are prepared to make a comparison with
numerical relativity.

\section{Head-on Collision of Spinning Black Holes with Transverse, 
Anti-parallel Spins}
\label{sec:implementation}

In this section, we discuss the specific example of a head-on collision
of equal-mass black holes with transverse, anti-parallel spins.
We will specialize the general framework presented in Sec.\ 
\ref{sec:qualitative} to the current configuration in the first subsection
and then make the comparison with numerical relativity in the second.

\subsection{The Hybrid Model for the Head-on Collision}
\label{sec:hybrid}

We will mimic the configuration used in the numerical-relativity simulation
for ease of comparison.
We thus choose our two black holes, labeled by $A$ and $B$, to have masses 
$M_A=M_B=M/2$, to start with initial separation $X_A=A(0)/2=-X_B$ 
[$A(0)=7.8M$ in the numerical simulations and $Y_A=Y_B=Z_A=Z_B=0$] and to have 
their spins along  $\pm Z$ axis, respectively (so that  $S_A^Z=0.5 M_A^2$ and 
$S_B^Z = -0.5M_B^2$ and all other components of the spins are zero).
Though they initially fall from rest, as in the numerical simulation,
we will denote their speeds by $V_A$ and $V_B$.

\subsubsection{Even-Parity Perturbations}

As we argued in Sec.\ \ref{sec:qualitative}, the even-parity perturbation
will only rely upon the Newtonian potential, which has the familiar form,
\begin{equation}
U_N = \frac{M_A}{R_A} + \frac{M_B}{R_B} \, .
\end{equation}
Here $R_A$ and $R_B$ denote the distance from the centers of black holes 
$A$ and $B$ in the PN coordinates. 
We then expand the Newtonian potential, $U_N$, into multipoles, keeping only 
the monopole and quadrupole pieces (the dipole piece vanishes),  
\begin{eqnarray}
&&U_N=U_N^{(0)} + U_N^{(2)}\\
\nonumber
&&=\frac{M}{R}+\frac{MA(t)^2}{4R^3} \sqrt{\frac{3\pi}{10}}
\left[ Y_{2,-2} -\sqrt{\frac{2}{3}}Y_{2,0} + Y_{2,2} \right]\,.\;
\end{eqnarray}
$Y_{l,m}$ are the usual scalar spherical harmonics.
One can then see that the nonzero coefficients of the spherical harmonics are 
\begin{equation}
U_N^{2,\pm 2} = \sqrt{\frac{3\pi}{10}}\frac{MA(t)^2}{4R^3} = -\sqrt{\frac{3}{2}}
U_N^{2,0} \, .
\end{equation}
After applying the transformation of the PN and Schwarzschild radial
coordinates, $R=r-M$, (and similarly $A(t) = a(t) - 2M$) one finds that
\begin{equation}
U_N^{2,\pm 2} = \sqrt{\frac{3\pi}{10}} \frac{M(a(t)-2M)^2}{4(r-M)^3} = 
-\sqrt{\frac{3}{2}}U_N^{2,0}  \, .
\end{equation}
One can then substitute this into Eq.\ (\ref{z_function_gen}) to find
the Zerilli function,
\begin{equation}
\Psi_{(\rm e)}^{2,\pm 2} = \sqrt{\frac{3\pi}{10}} \frac{Ma(t)^2}{2r^2} 
\left(1-\frac{7M}{6r}\right) = -\sqrt{\frac 32} \Psi_{(\rm e)}^{2,0}  \, .
\end{equation}
We have only kept terms to linear order in $M/r$ in this calculation,
because we only use Newtonian physics to calculate the gravitational
potential.
At the boundary of $r_s(t)=a(t)/2$, the perturbation is constant at leading 
order, and varying only at higher orders.
\begin{eqnarray}
\left.\Psi_{(\rm e)}^{2,\pm 2}\right|_{\rm shell} &=& \sqrt{\frac{6\pi}{5}} M
\left(1-\frac{7M}{3a(t)}\right)\,,\\
\left.\Psi_{(\rm e)}^{2,0}\right|_{\rm shell}&=&-\sqrt{\frac{4\pi}{5}} M
\left(1-\frac{7M}{3a(t)}\right)\, .
\label{even_bdry}
\end{eqnarray}

\subsubsection{Odd-Parity Perturbations}

The calculation with the gravitomagnetic potential is slightly more
difficult, because it involves additional gauge transformations.
The gravitomagnetic potential is given by
\begin{equation}
w_i = \frac{\epsilon_{ijk} S^j N_A^k}{2R_A^2} + \frac{M_AV_A^i}{4R_A}
+ \frac{\epsilon_{ijk} S^j N_B^k}{2R_B^2} + \frac{M_BV_B^i}{4R_B} \, .
\end{equation} 
These results appear, for example, in Eq.\ (6.1d) of \cite{Kaplan}.
The new symbols $N_A^k$ and $N_B^k$ represent unit vectors pointing from the 
centers of the two black holes in the PN coordinates, and $V_A^i$ and $V_B^i$
are the velocities of the two black holes.
Expanding the gravitomagnetic potential to leading order in $A(t)$ and 
simplifying the trigonometric portions of the equations below, we see that
\begin{eqnarray}
w_x &=&\frac{MVA(t)\sin\theta\cos\varphi}{2R^2}-
\frac{3A(t) S \sin^2\theta\sin2\varphi}{4R^3} \,, \\
w_y &=&-\frac{A(t)S(1+3\cos2\theta-6\sin^2\theta\cos2\varphi)}{8R^3} \,,\\
w_z&=&0 \,,
\end{eqnarray}
The variables $S$ and $V$ are just the magnitudes of the spin and velocity 
of each black hole, respectively.
For this equal-mass collision, the spins have the same magnitude, and the
velocities of the holes do, as well.
We must then convert the gravitomagnetic potential into spherical-polar 
coordinates,
\begin{eqnarray}
w_R &=& \frac{MVA(t)\sin^2\theta\cos^2\varphi}{2R^2}
-\frac{A(t)S\sin\theta\sin\varphi}{2R^3} \,,\\
w_{\theta} &=& \frac{MVA(t)\sin 2\theta\cos^2\varphi}{4R}
-\frac{A(t)S \cos\theta\sin\varphi}{2R^2}\,,\\
\nonumber
w_{\varphi}&=& -\frac{MVA(t)\sin^2\theta\sin 2\varphi}{4R}\\
&& +\frac{A(t)S\cos\varphi(5\sin\theta-3\sin3\theta)}{8R^2}  \, .
\end{eqnarray}
There is a dipole term in the component $w_R$ of the gravitomagnetic potential,
and this term will not evolve according to the Regge-Wheeler-Zerilli equation.
One can remove it via the small gauge transformation to the metric,
\begin{equation}
\hat h_{\alpha\beta} = h_{\alpha\beta}-\xi_{\alpha|\beta} -
\xi_{\beta|\alpha} \, , 
\end{equation}
where the bar refers to a covariant derivative with respect to the 
background metric (in this case flat space).
Recall that the metric components, $h_{ti}$, are related to the gravitomagnetic
potential, $w_i$, by $h_{ti} = -4w_i$.
If we make a gauge transformation where the only nonzero component 
of $\xi_\mu$ is
\begin{equation}
\xi_t=\frac{2MA(t)V\sin^2\theta\cos^2\varphi}{R}
-\frac{A(t)S \sin\theta\sin\varphi}{R^2} \, ,
\end{equation}
this has several important effects.
For one, it eliminates $\hat h_{tr}$, and it introduces a term, 
\begin{equation}
-\xi_{t|t} = -\frac{4M\sin^2\theta\cos^2\varphi}{R}
\left(V^2 - \frac{MA(t)}{2R^2}\right)  \, ,
\end{equation}
into $h_{tt}$.
This term, however, is of 1PN order, and, since we are considering only the
leading Newtonian physics, we will drop it.
Then, it turns the remaining perturbation into the sum of odd- and 
even-parity quadrupole perturbations.
Letting $b=\theta,\varphi$, one has that
\begin{eqnarray}
\hat h_{tb}&=&-\frac{A(t)S}{R^2}\sqrt{\frac{6\pi }{5}}(X^{2,1}_b-X^{2,-1}_b),\\
\nonumber
&& - \sqrt{\frac{2\pi}{15}}\frac{4MA(t)V}{R}\left(Y^{2,2}_b - \sqrt{\frac{2}{3}}
Y^{2,0}_b + Y^{2,-2}_b\right) \, .
\end{eqnarray}
If one were to include the even-parity, vector-harmonic term in the Zerilli 
function, one would need to take its time derivative.
This means it enters as a next-to-leading-order effect, and we can ignore that 
term in our leading-order treatment.
Thus, the relevant perturbation of the gravitomagnetic potential is
\begin{equation}
w_{(\rm o)}^{2,1} = \sqrt{\frac{6\pi }{5}}\frac{A(t) S}{4R^2} = 
-w_{(\rm o)}^{2,-1}\, .
\end{equation}
Finally, we make the transformation to the Schwarzschild radial coordinate,
$R=r-M$ (and similarly for $A(t) = a(t) - 2M$), to find that
\begin{equation}
w_{(\rm o)}^{2,1} = \sqrt{\frac{6\pi }{5}}\frac{(a(t)-2M)S}{4(r-M)^2} 
= -w_{(\rm o)}^{2,-1}\, .
\end{equation}
We can then find the  Regge-Wheeler function from Eq.\ (\ref{rw_function_gen}) 
which is 
\begin{equation}
\Psi_{(\rm o)}^{2,1}=-\sqrt{\frac{6\pi}{5}}\frac{2a(t)S}{r^2} 
= -\Psi_{(\rm o)}^{2,-1}\, .
\end{equation}
As before, we keep only terms linear in $M/r$.
At the boundary, the odd-parity perturbation is 
\begin{equation}
\left.\Psi_{(\rm o)}^{2,1}\right|_{\rm shell}=-\frac{8 S}{a(t)}
\sqrt{\frac{6\pi}{5}} = \left.-\Psi_{(\rm o)}^{2,-1}\right|_{\rm shell}\, .
\label{odd_bdry}
\end{equation}

\subsubsection{Energy and Momentum Fluxes}
Finally, because we only have quadrupole perturbations, the expressions
for the energy and momentum fluxes greatly simplify.
The energy flux, for the $l=2$ modes (taking into account that the $m=\pm1$
are equal and the $m=\pm2$ modes are equal, as well) becomes
\begin{equation}
\dot{E} = \frac{3}{8\pi}\left[\frac 83 \left(\dot{\Psi}^{2,2}_{(\rm e)}
\right)^2+2\left(\dot{\Psi}^{2,1}_{(\rm o)}\right)^2\right] \, ,
\end{equation}
and the momentum flux is given by
\begin{equation}
\dot{P}_y = \frac{1}{\pi} \dot{\Psi}_{(\rm e)}^{2,2} \dot{\Psi}_{(\rm o)}^{2,1} \, .
\label{kick_py}
\end{equation}
We have also used the fact that
$\Psi_{(\rm e)}^{2,0} = \sqrt{2/3}\Psi_{(\rm e)}^{2,\pm2}$ in this head-on
collision.

\subsection{Comparison with Numerical Relativity}
\label{sec:comparison}

In this section we compare the results of our head-on collision of spinning
black holes (with transverse, anti-parallel spins) with the equivalent
results from a numerical-relativity simulation (see the paper by Lovelace
et al.\ \cite{Lovelace} for a complete description of the simulation).
Although the paper by Lovelace et al.\ dealt mostly with
using the Landau-Lifshitz pseudotensor to define a gauge-dependent 4-momentum 
and an effective velocity to help develop intuitive understanding of 
black-hole collisions, they also investigated the gauge-invariant
gravitational waveforms and radiated energy-momentum (calculated
from the gravitational waves at large radii).
We will not attempt to study any of these Landau-Lifshitz quantities
in this work, and, instead, we will just look into the gauge-invariant
radiated quantities.
Specifically, for our comparison, we focus on the waveforms (the $l=2$ modes 
of the gravitational waves) and the radiated energy and momentum.

\begin{figure}[htb]
\includegraphics[width=0.45\textwidth]{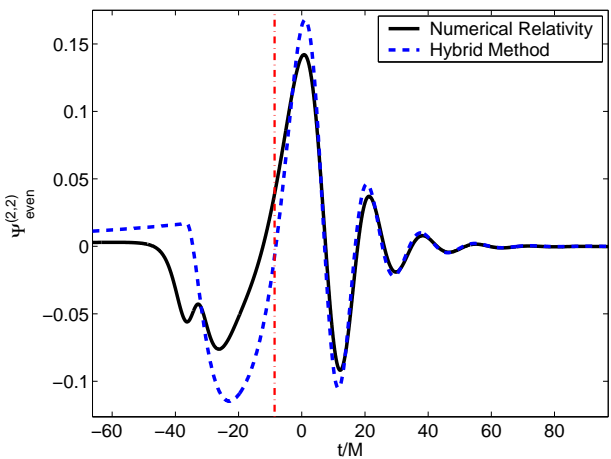}
\caption{(Color online) The blue (dark gray) dashed curve is
$\Psi_{\rm even}^{2,2}$ from our hybrid method, whereas the black solid curve
is the same quantity in full numerical relativity.
The red (light gray) dashed vertical line corresponds to the retarded time at
which the shell in the hybrid method reaches the light ring of the final
black hole, $r=3M$.
We shift the numerical-relativity waveform so that the peaks of the numerical
and hybrid $\Psi_{\rm even}^{2,2}$ waveforms align.}
\label{even_wave}
\end{figure}

\begin{figure}[htb]
\includegraphics[width=0.45\textwidth]{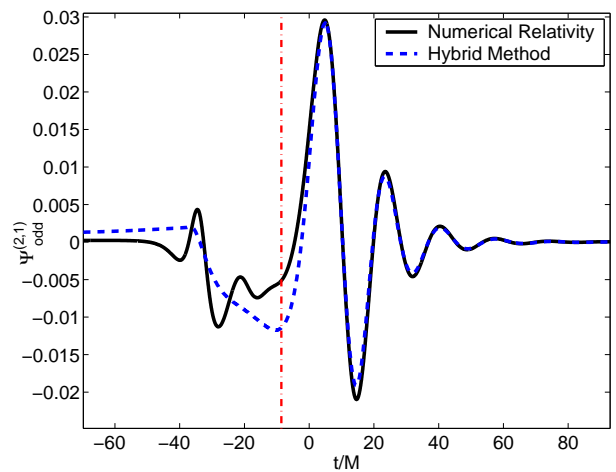}
\caption{(Color online) The blue (dark gray) dashed curve is
$\Psi_{\rm odd}^{(2,1)}$ from our hybrid method, whereas the black solid
curve is the same quantity in full numerical relativity.
The red (light gray) dashed vertical line corresponds to the retarded time at
which the shell in the hybrid method reaches the light ring of the final
black hole, $r=3M$.
We shift the numerical-relativity waveform so that the peaks of the numerical
and hybrid $\Psi_{\rm even}^{2,2}$ waveforms align.}
\label{odd_wave}
\end{figure}

In Figs.\ \ref{even_wave} and \ref{odd_wave}, we compare, respectively,
the even-parity perturbation $\Psi_{\rm even}^{(2,2)}$  and the odd-parity 
perturbation $\Psi_{\rm odd}^{(2,1)}$ from our method with the equivalent
quantities from the numerical simulation S1 featured in Lovelace et al.
(Since the $l=2$, $m=-2,-1,0$ components are related by
constant multiples to the above perturbations, we do not show them.)
In these figures, we also include a vertical, dashed red line that indicates
the retarded time at which the shell crossed the light-ring of the final
black hole $r=3M$, in the hybrid method.
This is the peak of the effective potential, and due to the low-frequency
opacity of this potential, much of the influence of the boundary data is
hidden within the potential after this time (and the waveform is due mostly
to the quasinormal modes of the final black hole).
Before this time the match is not exact (as a result of junk radiation in the
numerical simulation and the difference between the time coordinates), but
the Newtonian order perturbations do quite a good job of exciting quasinormal
modes of a reasonable amplitude.

The even- and odd-parity waveforms in the hybrid method are the pieces of
$\Psi_{\rm even}^{(2,2)}$ and $\Psi_{\rm odd}^{(2,1)}$ restricted to the
outer boundary of the characteristic grid, labeled by {\it BC} in Fig.\
\ref{fig1}.
We found these perturbations through the procedure described in Sec.\ 
\ref{sec:details}, applied to the specific binary parameters described
in Sec.\ \ref{sec:hybrid}.
For the numerical-relativity waveforms, we chose to present them in 
terms of the even- and odd-parity perturbation functions 
$\Psi_{\rm even}^{(2,2)}$ and $\Psi_{\rm odd}^{(2,1)}$, as well.
To find these perturbation functions from the numerical simulations, we 
first twice integrated the Weyl scalar $\Psi_4$ with respect to time to
get the waveforms $h_+$ and $h_\times$ 
(since $-\Psi_4 = \ddot h_+ -i \ddot h_\times$, at large radii, where a dot 
denotes a time derivative).
One can relate them to the gravitational waveforms $h_+$ and $h_\times$ by
Eq.\ (\ref{h_plus}), at large $r$.
In the case of the $l=2$ perturbations shown here,
$r h_+^{(2,2)} = \sqrt{6} \Psi_{\rm even}^{(2,2)}$ and 
$r h_\times^{(2,1)} = \sqrt{6} \Psi_{\rm odd}^{(2,1)}$.
We compared the $h_+$ and $h_\times$ found directly from the numerical
simulation through extraction of the Regge-Wheeler and Zerilli functions from
metric coefficients in the numerical code \cite{Rinne}, and the two procedures 
gave essentially identical results.

In order to make the comparison between our hybrid method and the
full numerical-relativity waveforms, we must shift the numerical
waveforms by a constant. 
Specifically, we choose this constant so that the peaks of the exact and
approximate waveforms of $\Psi_{\rm even}^{(2,2)}$ match (at a time that
we set to be $t=0$).
We add this constant shift in time, because there is no
clear relationship between the coordinate time at which the waveform in our 
code begins and the same coordinate time in the numerical-relativity 
simulation.
Trying to find a relationship between these times is complicated by the fact
that the hybrid method evolves on a characteristic grid, whereas the 
numerical-relativity simulation solves an initial-value problem in a gauge 
that changes as the black holes move together.
Nevertheless, because both the numerical and the hybrid method use 
asymptotically flat coordinates, at large radii, the time coordinates
move at the same rate.
This, in turn, means that it is only necessary to shift the time coordinates
rather than rescaling them.
As an interesting aside, Owen \cite{Owen} found that this agreement between 
the time coordinates in the numerical simulation and perturbation theory 
appears to extend even into the near zone, when he observed that multipole
moments of the horizon oscillate at the quasinormal mode frequencies of the
black hole.
See the end of Sec.\ III of that paper for a discussion of why that might
be the case.

\begin{figure}[htb]
\includegraphics[width=0.45\textwidth]{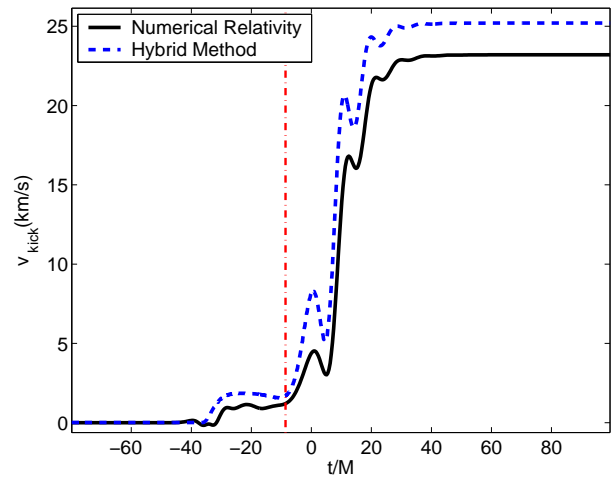}
\caption{(Color online) The blue (dark gray) dashed curve is the velocity of 
the final black hole as a function of time (inferred from the gravitational 
radiation) from our hybrid method, using only the $l=2$ modes of the wave.
The black solid curve is the equivalent quantity in numerical relativity, 
computed from the full Weyl scalar, $\psi_4$.
The red (light gray) dashed vertical line corresponds to the retarded time at 
which the shell in the hybrid method reaches the light ring of the final black 
hole, $r=3M$.
As before, the numerical-relativity velocity is shifted so that the peaks of 
the numerical and hybrid $\Psi_{\rm even}^{2,2}$ waveforms align.}
\label{kick}
\end{figure}

We also compute the momentum flux, and we show the accumulation of the 
velocity of the final black hole in Fig.\ \ref{kick}, for both our method
and the full numerical-relativity simulation.
For our hybrid method, we use just the $l=2$ modes of the 
waveform to compute the momentum flux, $\dot P_y$, our Eq.\ (\ref{kick_py}).
We then find the velocity of the final black hole as a function of time by
computing
\begin{equation}
v_y(t) = -\frac 1 M \int_{t_0}^t \dot{P}_y(t') dt' \, ,
\label{v_y}
\end{equation}
where we introduce an extra minus sign to account for the fact that the 
black-hole's velocity is opposite that of the momentum carried by the
gravitational waves.
For the numerical waveform, we show the equivalent velocity computed from 
the full Weyl scalar, $\Psi_4$.
For the numerical simulations, one typically computes
\begin{equation}
\dot{P}_y = \lim_{r\rightarrow\infty} \frac{r^2}{16\pi} \oint
\sin\theta\sin\phi \left | \int_{-\infty}^t \Psi_4 dt' \right |^2 d\Omega,
\end{equation}
where $\Psi_4$ is the Weyl scalar extrapolated to infinity, and $d\Omega$ is 
the surface-area element on a unit sphere.
This expression appears in a variety of sources [see, for example, Eq.\ (29) 
of the paper by Ruiz et al.].
We then can compute the velocity of the final black hole in the numerical
simulations through Eq.\ (\ref{v_y}), as we did for the hybrid method.
Again, we perform the same time-shifting procedure as we did with the 
waveforms.
The kick we find is remarkably close; 22 km/s for the numerical simulation
and 25 km/s for our hybrid method.

The radiated energy does not agree quite as well due to the fact that
the even-parity perturbation is somewhat larger than the equivalent
numerical quantity (and it is the dominant contribution to the energy
flux).
Nevertheless, the results agree within a factor of two; the numerical
simulation shows that roughly $0.057\%$ of the initial mass is radiated, 
whereas our hybrid method predicts approximately $0.096\%$ of the initial
mass escapes in gravitational waves.
This is an improvement over some of the earlier, first-order, 
close-limit calculations at larger separations (see, for example, Fig.\ 1 
of \cite{Gleiser}, where more than $1\%$ of the initial mass is radiated for
equivalent initial separations of the black holes).

We are not aware of any equivalent calculations of radiated 
energy for head-on collisions within the EOB formalism.
While there have been recent 3PN calculations of the energy flux for a head-on 
plunge of black holes by Mishra and Iyer \cite{Mishra}, in PN-based 
calculations such as the EOB approach, one must always stop the calculation
at some radial separation of the binary.
For example, in Fig.\ 2 of Mishra and Iyer's work, they stop the calculation 
at a point at which only roughly a tenth of the total energy that will be
radiated in the head-on collision, has escaped.
This poses a small problem for EOB approaches, because as described in the 
introduction to this paper, Sec.\ \ref{sec:introduction}, one must choose a 
point at which to stop the EOB inspiral-and-plunge waveform and match it
to a set of quasinormal modes to obtain a full waveform.
For inspiraling black holes, there is a natural point to do this: when the
frequency of the inspiral-plunge dynamics approaches the quasinormal mode
frequencies of the final black hole that will be formed.
For a head-on collision, however, there is no analogous frequency at which
one can match.
We will, therefore, reserve any comparisons between our method and 
that of EOB for future work, when we extend our method to inspiraling,
black-hole binaries.

\section{The Three Stages of Black-Hole Mergers}
\label{sec:discussion}

In addition to producing reasonably accurate full waveforms, our approach
also provides a possible interpretation of the infall, merger and 
ringdown stages of a binary-black-hole  merger.  
As shown in Fig.\ \ref{fig1}, before the shell reaches point {\it e} and 
enters the strong-field region [the red (dark gray) area, in which the $l=2$, 
even-parity, effective potential exceeds $1/3$ of its peak value], the majority 
of the retarded waveform propagates along the light cone (the so-called direct 
part). 
The direct part overwhelms the waves that scatter off the background
curvature, because, far away from the source, the curvature is small.
This part of the waveform corresponds to the inspiral or infall phase.

In the strong-field region, however, there is strong curvature (the 
black-hole effective potential in our model).
While some fraction of the waves will pass directly through, as the shell enters
this region at point {\it e}, waves that scatter off the effective potential 
(and thereby propagate within the light cone) become more significant.
These waves often are called tail waves.
Although PN waveforms do include the tail part, the fact that the higher-order
PN terms that contain the tail dominate over the lower-order terms 
\cite{Racine} does not bode well for the ability of the PN series to
easily capture this effect.
Nevertheless, we are able to associate this mixture of direct and tail
portions in the waveform to merger.
In our model, this stretch of the waveform is related to the retarded times 
when the shell is passing through the strong-field region of spacetime (points 
{\it e}, {\it f}, and {\it g} in the diagram).

Finally, after the shell passes through the potential, the details of the 
perturbation no longer become important, as was found by Price in his stellar
collapse model.
Because waves do not efficiently propagate through the barrier, the 
gravitational waveform associated with points {\it g} through {\it D}
should arise from before and while the shell passes through the effective
potential (not after).
This last piece is that of ringdown.  

There is one subtlety to note about our interpretation of ringdown that might
arise if the final black hole is a Kerr black hole. 
Mino and Brink \cite{Mino} and Zimmerman and Chen \cite{Zimmerman} showed that
mergers that lead to a Kerr black hole can emit waves at integer 
multiples of the horizon frequency that decay at a rate proportional to the 
surface gravity.
These modes come from a calculation in the near-horizon limit of a Kerr black
hole, and from the vantage point of observers far away, these waves would 
appear to be coming from the horizon.
These modes have a sufficiently high frequency that they could penetrate the 
effective potential of a Kerr black hole and contribute to the ringdown portion
of the gravitational wave.
Nevertheless, if we expand our description of the ringdown phase to include
these horizon modes, our interpretation holds more or less as described above.

\section{Conclusion}
\label{sec:conclusion}

In this paper, we show, by examining the head-on collisions of spinning 
black holes, that a combination of PN and BHP theories gives a gravitational 
waveform that matches well with that of full numerical-relativity 
simulations.
We were able to do this not by applying the approximation methods to distinct
times in the evolution of the system, but by choosing regions of space in which
the two methods work and finding that the waveform from black-hole-binary 
collisions can be protected from lack of convergence in these approximations.
Specifically, our method lumps all monopole pieces of a PN black-hole-binary
spacetime into the Schwarzschild metric and treats the higher multipoles as
perturbations of that Schwarzschild that evolve via a wave equation.
Moreover, since PN and BHP theories describe the waveform, this suggests that 
much of the nonlinear dynamics appearing in the gravitational waves of a 
head-on black-hole-binary merger can be well approximated by linear 
perturbations of the Schwarzschild solution.

Our approach certainly cannot replace full numerical simulations. 
For one, we must test its validity for different kinds of coalescence by 
comparison with fully nonlinear numerical results.
Nevertheless, we are hopeful that our method maybe be useful for gaining further
understanding of the spacetime of black-hole-binary mergers and for 
producing templates of gravitational waveforms for data analysis.
To move towards these goals, we would need to make several modifications to 
our method (whose implementation we leave for future work).
Most of these changes revolve around finding a way to treat inspirals of 
black-holes binaries within our method.
The most necessary addition would be finding a way to consistently treat 
radiation reaction within the formalism.
This feature is essential for capturing the correct inspiral and plunge
dynamics.
Also important for describing realistic physics of the ringdown would be
to analyze the problem in a Kerr background.
Each of these problems requires significant work, so we leave them for
future studies.

\acknowledgments

We thank Geoffrey Lovelace and Uli Sperhake for supplying waveforms and 
energy-momentum fluxes from their numerical simulations; we thank Lee 
Lindblom, Mark Scheel and B\'ela Szil\'agyi for advice on 
solving wave equations with characteristic methods.
We thank Drew Keppel for his input in discussions during the early stage of
this work, and we thank Kip S.\ Thorne and Yasushi Mino for discussing
related aspects of black-hole physics with us. 
This work has been supported by NSF Grants No.\ PHY-0601459, PHY-0653653 and 
CAREER Grant PHY-0956189, by the David and Barbara Groce start-up funds at the 
California Institute of Technology, and by the Brinson Foundation.
D.N.'s research was supported by the David and Barbara Groce Graduate Research
Assistantship at the California Institute of Technology.

\end{document}